\documentstyle[11pt,newpasp,twoside,epsf]{article}
\def\edcomment#1{\iffalse\marginpar{\raggedright\sl#1\/}\else\relax\fi}
\reversemarginpar
\input{epsf}
\input{rotate}

\begin{document}
\title{The HI Content and Extent of Low Surface Brightness Galaxies -- 
Could LSB Galaxies be Responsible for Damped Ly-$\alpha$ Absorption?}
\author{Karen O'Neil}
\affil{NAIC/Arecibo Observatory, HC3 Box 53995, Arecibo, PR 00612}
\affil{koneil@naic.edu}


\begin{abstract}

Low surface brightness galaxies, those galaxies with a central surface brightness
at least one magnitude fainter than the night sky, are often not included in 
discussions of extragalactic gas at z $<$ 0.1.  In this paper we review many
of the properties of low surface brightness galaxies, including recent studies
which indicate low surface brightness systems may contribute far more to
the local HI luminosity function than previously thought.  Additionally, we
use the known (HI) gas properties of low surface brightness galaxies to
consider their possible contribution to nearby damped Lyman-$\alpha$ absorbers.
\end{abstract}

\section{Introduction - What is a LSB Galaxy?}

Typically, low surface brightness (LSB) galaxies are defined as those galaxies with an
{\it observed} central surface brightness that is at least one magnitude fainter
than the night sky.  In the B band, this translates to $\mu_B(0)\:\ge$ 22.6 -- 23.0
mag arcsec$^{-2}$.  However, alternatives to this definition do exist.  One
common definition is that a LSB galaxy is a galaxy whose {\it inclination
corrected} central surface brightness is  $\mu_B(0)\:\ge$ 23.0
mag arcsec$^{-2}$ (Matthews, Gallagher, \& van Driel 1999). Although
it is a more consistent definition, this second definition relies
on understanding the dust properties and opacity of the studied galaxies.  As
detailed studies of the dust content of the majority of the LSB galaxies 
discussed herein have not been done, we will restrict the definition
of LSB systems in this paper to only those galaxies with an observed $\mu_B(0)\:\ge$ 23.0
mag arcsec$^{-2}$.

Photometrically, LSB galaxies are typically thought of as being fairly blue ($B-V\:<\:0.3$)
and unevolved.  Although it is true LSB galaxies include perhaps the bluest galaxies
known (e.g. O'Neil, et al. 1998; de Blok, van der Hulst, \& Bothun 1996; McGaugh, Schombert, \& Bothun 1995),
a number of LSB systems with $B-V\:>$ 1
have also been identified (O'Neil, et al. 1997b).  Whether the red LSB systems are
indicators of a bias towards detecting blue LSB systems (O'Neil, et al.. 1997b),
examples of an uncommon quiescent phase of star formation in LSB galaxies (Gerristson \&
de Blok 2000), or LSB galaxies with a different star formation history than their
blue counterparts (Bell, et al. 2000), or some combination of the three, remains to be seen.
The important point is simply that LSB systems are not exclusively blue galaxies,
and allowances for LSB galaxies with redder colors must be made.

The last common descriptor of LSB galaxies is their morphology.  Typically, LSB galaxies
are described as having the morphology of late-type/irregular spiral galaxies.  This description
certainly holds true for a significant percentage of LSB systems, many of which have
morphologies which simply cannot be described by the Hubble diagram (Figure 1).
LSB galaxies are not, though, exclusively of irregular  morphology -- a significant number
of LSB systems exist which have diffuse yet well defined outer disks and occasionally
central bulges.  Additionally, there are a number of LSB dE systems which have been identified.
(See, e.g. Impey, Bothun, \& Malin 1988; Evans, Davies, \& Phillipps 1990; Pickering, et al. 1997;
Bijersbergen, de Blok, \& van der Hulst 1999).

\begin{figure}[ht]
\plottwo{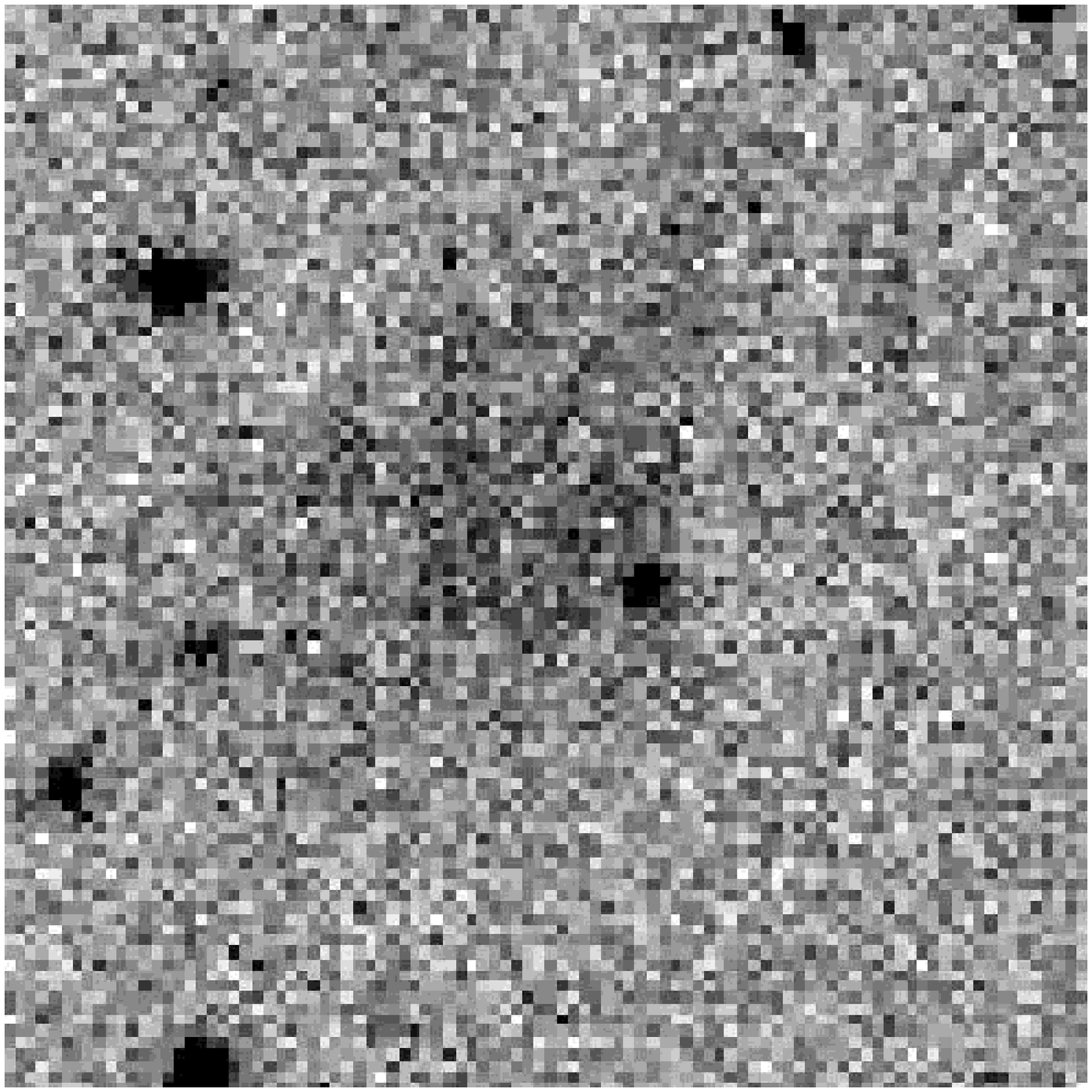}{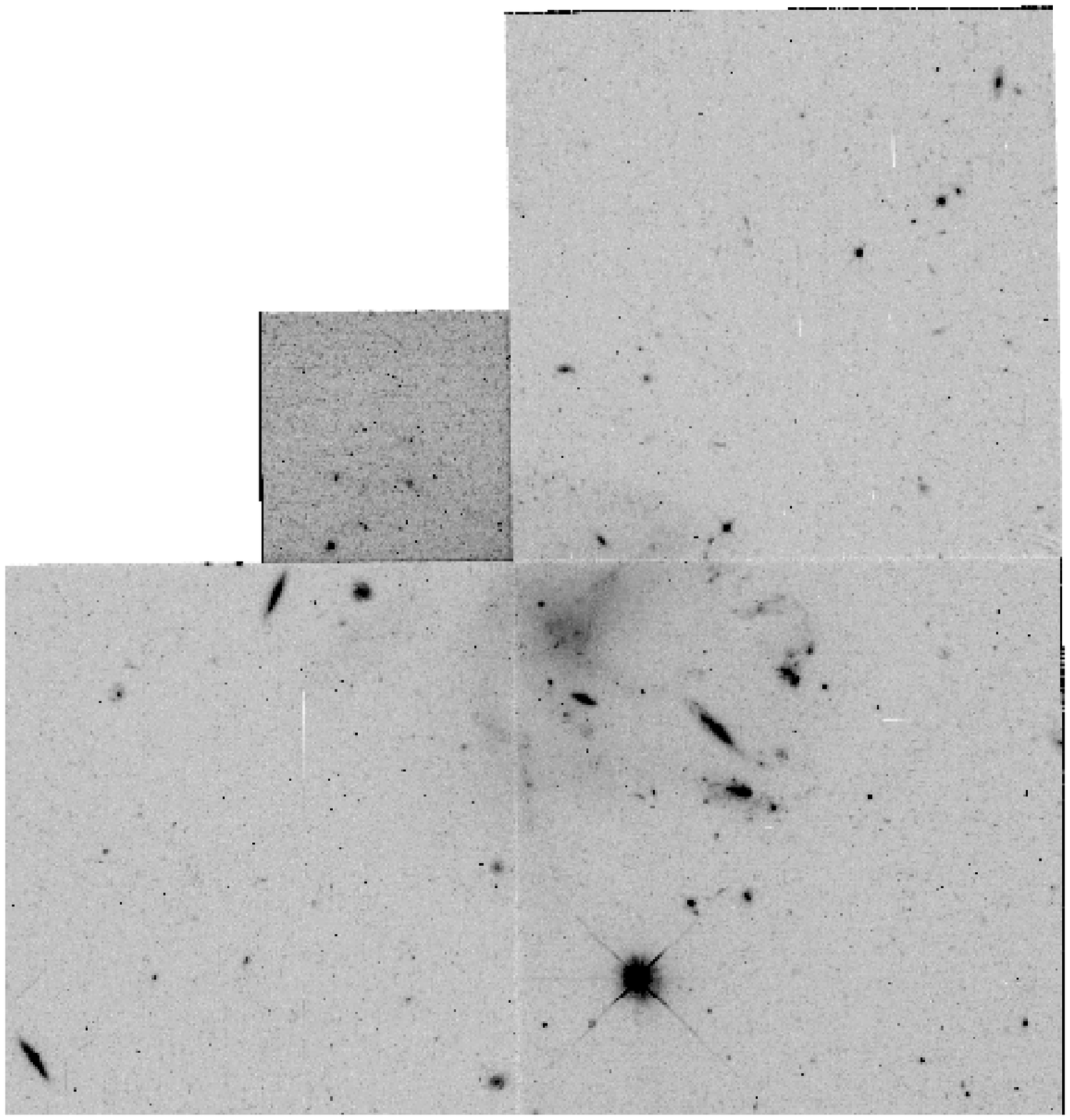}
\vskip -1.in
\caption{Images of [OBC97] P02-4 (left) and UGC 12695 (right)
-- LSB galaxies with $\mu_B(0)$ = 25.1 and 23.8 mag arcsec$^{-2}$, 
respectively.  (Images from O'Neil, Bothun, \& Cornell 1997a and O'Neil, McGaugh, \& Verheijen 1999.)\label{fig:morph}}
\end{figure}

\section{What is the HI Content of LSB Galaxies?}

Like many of their properties, the total neutral hydrogen content of
LSB galaxies varies considerably, from less than 10$^8$M$_\odot$ through
10$^{11}$M$_\odot$.  Contrary to often held belief, though, no correlation
is seen between the galaxies' total gas mass and the galaxies' central 
surface brightness or size (Figure 2a).  
Thus LSB galaxies are neither exclusively dwarf nor massive systems, but
instead cover the same mass range as their HSB counterparts.

Similarly, the gas mass-to-luminosity ratio of LSB galaxies varies
considerably, from less 0.1 $\le\:M_{HI}/L_B\:\le$
10 M$_\odot$/L$_\odot$.  Again, no correlation is seen between
LSB galaxies' $M_{HI}/L_B$ and surface brightness or size (Figure 2b),
indicating LSB galaxies are not exclusively of high (or low) gas
content.

\begin{figure}[h]
\plottwo{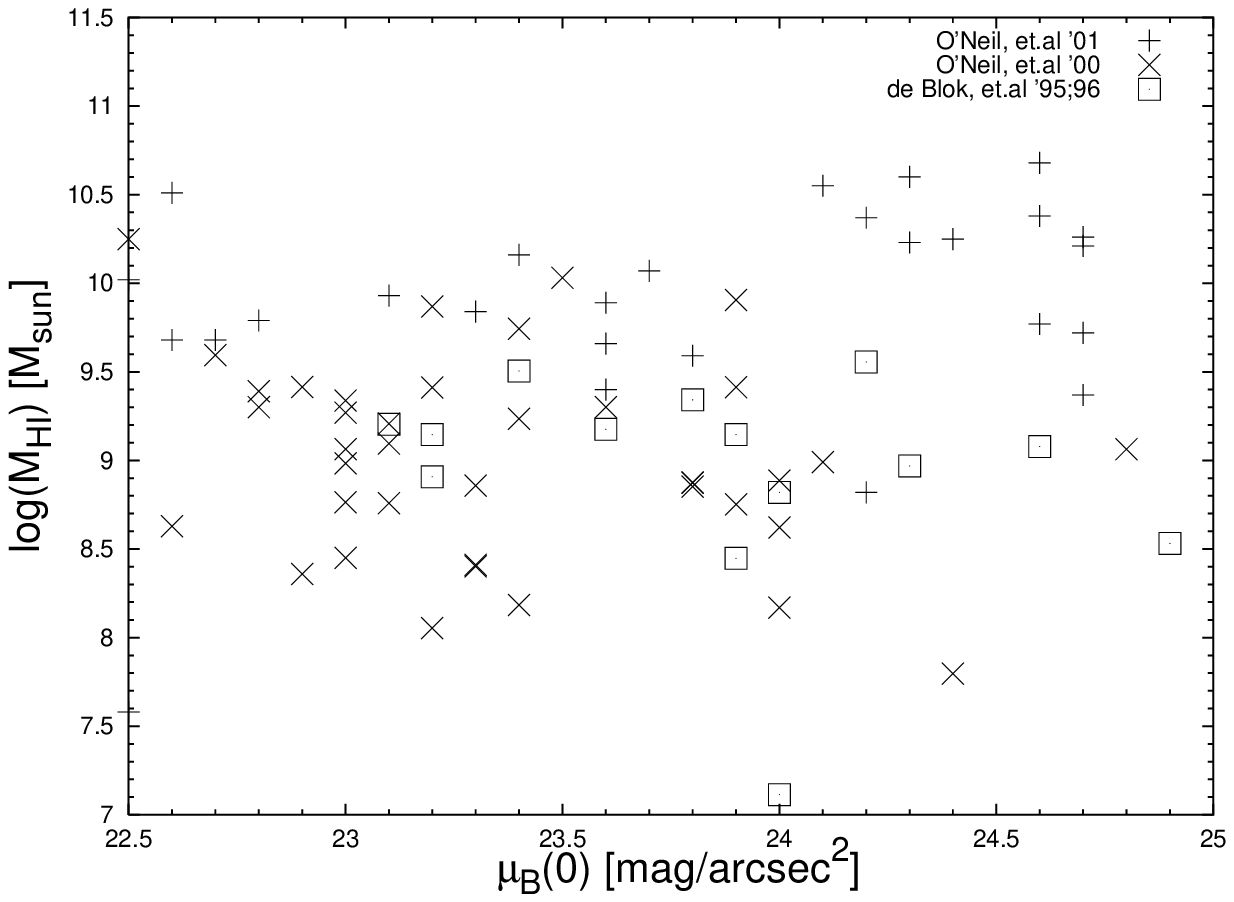}{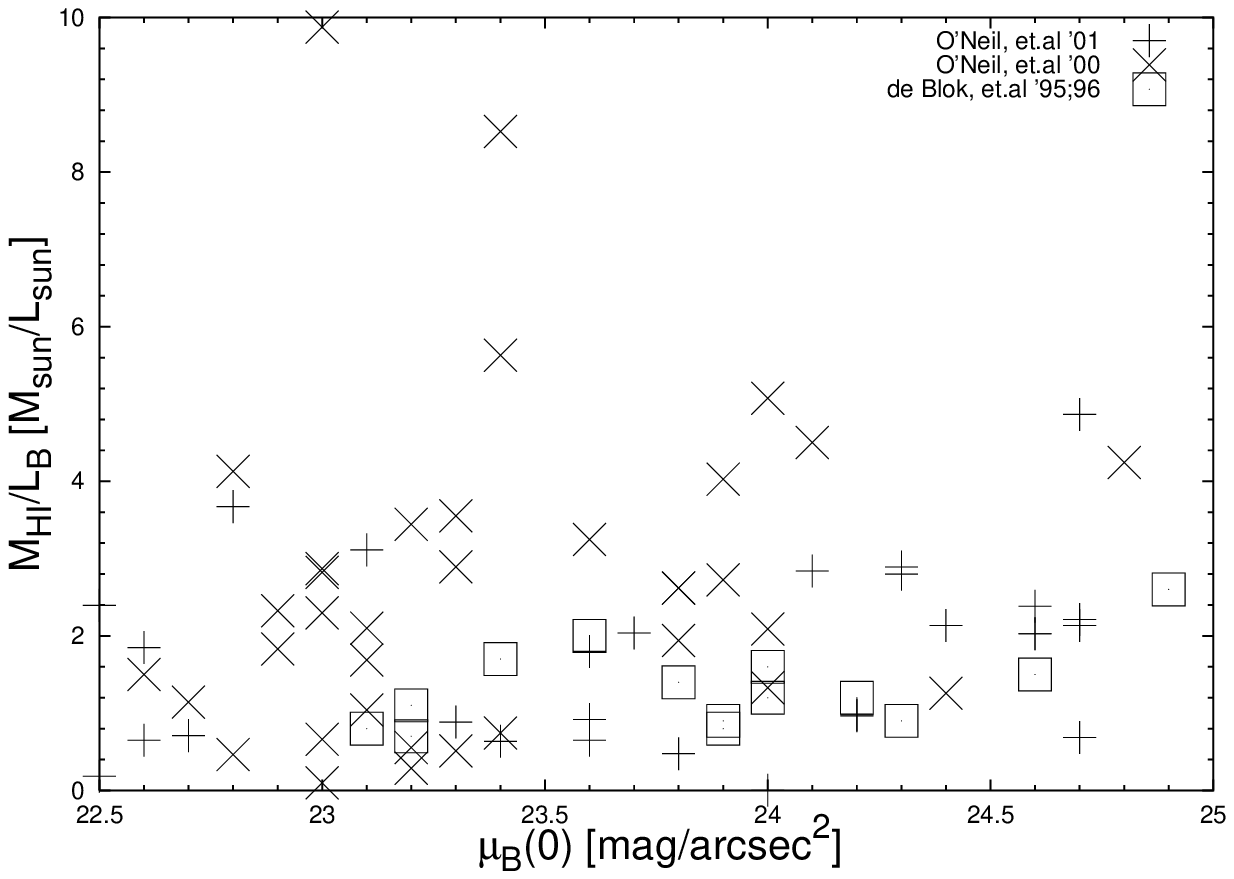}
\caption{Gas mass (a) and gas mass-to-luminosity ratio (b) versus central surface brightness 
for LSB galaxies in a variety of surveys. \label{fig:himass}}
\end{figure}

Recently, the number of massive (M$_{HI}$ $>$ 10$^{10}$ M$_\odot$) LSB galaxies known
has increased significantly.  Two survey are underway using the 305-m telescope at Arecibo
to determine the redshift and HI mass of optically selected LSB galaxies.
The first survey is searching for LSB galaxies in the UGC catalog which
(a) have not yet had a redshift determination, and (b) are morphologically
similar to the known massive LSB galaxies (e.g. Malin 1 and its `cousins')
(O'Neil \& Bothun 2001).  The galaxies from the second survey have been identified
(for the first time) on the POSS II plates, and are morphologically similar
to the known LSB AGN galaxies (Schombert, O'Neil, \& Eder 2002; Schombert 1998).

Although both massive LSB galaxy surveys are still underway,
preliminary results are available.
First, early results show that the surveys have increased the number of known massive
LSB galaxies by a factor of seven or more.   Additionally, the surveys have
significantly increased the number of LSBG AGN galaxies.
With this data in hand, it becomes evident that LSB galaxies may
contribute significantly to the HI luminosity function, in contradiction
to the commonly held belief of LSB systems as insignificant contributors 
to this function (i.e. Zwaan, these proceedings).  This argument is furthered by the 
discovery of numerous massive LSB systems with the HIPASS equatorial survey
(Disney, et al. in these proceedings).
Additionally, the findings of the two massive LSB galaxy
surveys have considerably increased the number of LSB galaxies with high
($>$200 kpc) impact parameters, a finding which plays a significant 
role in determining the contribution of LSB systems to damped Lyman-$\alpha$
systems (below).
Finally, and perhaps most importantly, the surveys' results
demonstrate the need for a better understanding
of HI survey selection effects. (A fact which is again further emphasized when the
results are combined with the HIPASS equatorial survey results presented 
by Disney (these proceedings).)

Although it now seems clear that there are a number of massive LSB galaxies in the
z $<$ 0.1 Universe, some caution is necessary.  First, unlike the HIPASS survey,
the two massive LSB galaxy surveys described herein are extremely biased.
As the surveys are designed to find massive LSB systems, the surveys' findings
cannot be used to directly re-determine LSB galaxies' contribution to 
the local HI luminosity function.  Instead the survey results should be
used simply as an indication that a significant population of massive LSB
galaxies do exist which have not yet been cataloged.

Secondly, it should be recognized that the
majority of LSB galaxy redshifts have been determined through observing the galaxies'
21-cm line.  This means that a galaxy is typically included in HI catalogs only if it
has a significant HI flux, as without a redshift known a priori to the 21-cm observations,
now meaningful upper limit to the galaxies' HI mass can be made.  As LSB galaxies are 
inherently low signal-to-noise systems, this limitation results
in significant selection effects being placed on all catalogs of LSB galaxy HI measurements
(Figure 3).  These selection effects, then, may be the primary reason LSB galaxies are 
typically considered gas rich objects.
 
\begin{figure}[ht]
\plotfiddle{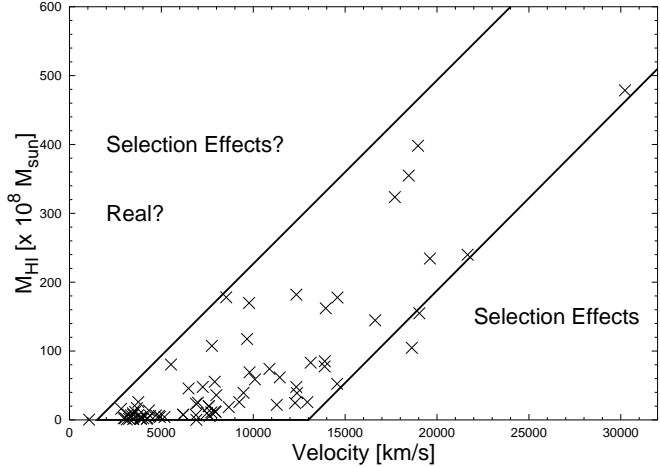}{2.1in}{0}{70}{70}{-170}{-40}
\caption{Total HI gas mass versus recessional velocity for the LSB galaxies in O'Neil, et al.
(2000) and O'Neil \& Bothun (2001).}
\end{figure}

\section{The HI Distribution of LSB Galaxies}

The only study done to date specifically looking at the local environment
of LSB galaxies was done by Bothun, et al. (1993).  
Using a sample of 340 LSB galaxies with measured redshifts and the CfA redshift survey data,
Bothun, et al. determined that LSB galaxies have a strong statistical
deficit of galaxies located within a projected radius of 0.5 Mpc and a velocity of 500 km/s
compared to HSB galaxies.  Comparing LSB and HSB disk galaxies in the same
portion of the sky, they found the average distance to the nearest neighbor is 1.7 times farther for LSB disk
galaxies.
On larger scales, though, LSB galaxies lie in the same overall distribution as HSB galaxies,
and no evidence has yet been found for LSB galaxies existing inside the large-scale
galaxy voids (Figure 4).

\begin{figure}[ht]
\plotfiddle{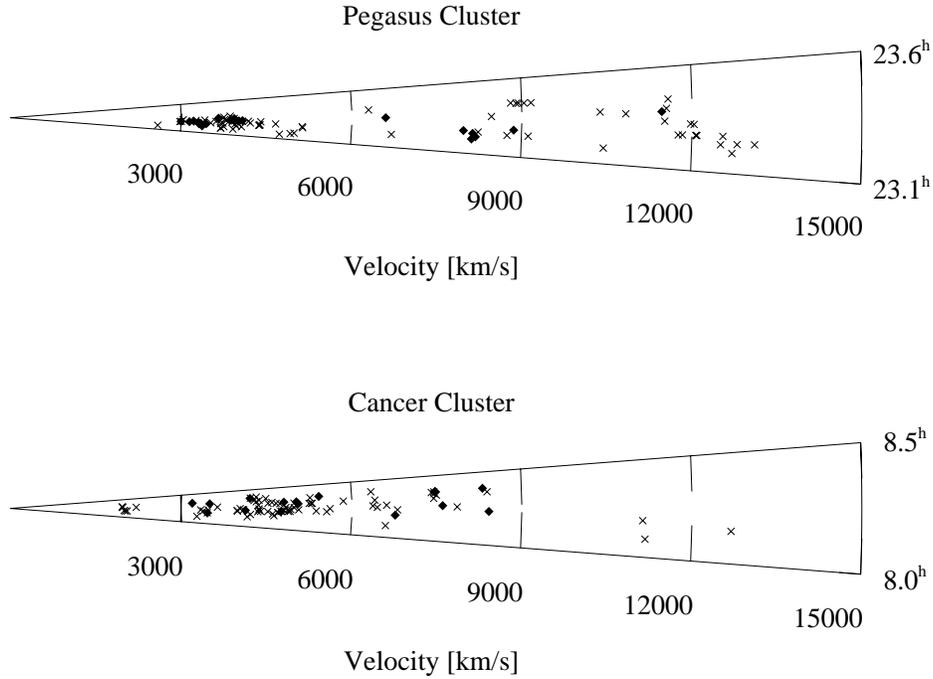}{4.0in}{90}{70}{70}{220}{-170}
\caption{Two dimensional projection of LSB galaxies in the O'Neil, Bothun, \& Schombert (2000)
survey ($\bullet$) as well as all other HSB galaxies with published velocities and lying in
the same sky region (x) (as determined from
NED).  The plots show heliocentric velocity (in km/s) versus Right Ascension (B1950 coordinates).}
\end{figure}

\section{The HI Extent of LSB Galaxies}

The optical radii (r$_{27}$) of LSB galaxies range from 
1 kpc through greater than 100 kpc, while the HI distribution
of LSB systems is typically much larger.
This difference was quantified be de Blok, et al. (1996).
Using the VLA and Westerbock
Synthesis Radio telescope, de Blok, et al. imaged the 21-cm line
of 19 LSB galaxies.  Defining the HI radius as the radius when the
HI distribution falls to the 
1 M$_\odot$ pc$^{-2}$ HI level, they determined
the HI to optical radius ratio of the studied LSB systems.
The results of this study found the R$_{HI}$/R$_{25}$
ratio to vary from 1.0 -- 4.3, with $\langle R_{HI}/R_{25} \rangle$ = 2.5,
showing LSB galaxies to have extended HI, potentially out to radii
of 200--300 kpc or more at the 10$^{18}$ cm$^{-2}$ level.

\section{Could LSB Galaxies be Significant Contributors to 
Damped \\{Lyman-$\alpha$} Absorption?}

One of the more significant questions which can be asked when considering
the gas content and number density of LSB galaxies is -- 
Could LSB galaxies be significant contributors to
damped Lyman-$\alpha$ absorption?
The traditional answer to this question has been no, for a variety of
reasons.  Namely, 
\begin{itemize}
\item It is often argued that LSB galaxies' HI column density is too low
for LSB galaxies to be damped Lyman-$\alpha$ absorbers, at least at 
large (100+ kpc) impact parameters (e.g. Zwaan, Verheijen, \& Briggs 1999)  
\item LSB galaxies are not found when catalogs are examined to determine what galaxies may lie within
the line-of-sight (e.g. Rao \& Turnshek 1998; Chen, et al. 1998)
\item There have been a number of searches undertaken to identify known damped Lyman-$\alpha$ absorbers with 
LSB galaxies which have failed (e.g. Rauch, Weyman, \& Morris 1996)
\end{itemize}

Recently, though, a variety of observations have been obtain which contradict
the above arguments, and which can be used to argue that LSB systems
are likely very important Lyman-$\alpha$ absorbers.
First, as was discussed in the last section, there appears to be a large number of
massive LSB galaxies with ample gas at high (greater than 100 kpc) impact parameters.
Additionally, there are clearly many LSB systems which have yet to 
be cataloged.  And although many of these galaxies may not have extended
high density gas, the gas density at the center of these systems
is more than sufficient to account for the gas density observed in
damped Lyman-$\alpha$ systems.  
Finally, and perhaps most importantly, it should be pointed out that a number
of recent observational studies have successfully identified LSB
galaxies as the most likely producers of damped Lyman-$\alpha$ absorption
(e.g. Bowen, Tripp, \& Jenkins 2001; Steidel, et al. 1994; Turnshek, et al. 2001).
As more observations are made, then, the possibility that LSB galaxies are
significant contributors to damped Lyman-$\alpha$ absorption continues to increase.

\section{Conclusion}

As has been discussed herein, the neutral hydrogen content
of LSB galaxies covers a wide range.  The total (HI) gas content
of LSB galaxies range from less than 10$^8$ M$_\odot$ through
10$^{11}$ $M_\odot$, while the M$_{HI}$/L$_B$ ratio
ranged from less than 1 M$_\odot$/L$_\odot$ through greater
than 10 M$_\odot$/L$_\odot$ (and possibly even greater than 100 M$_\odot$/L$_\odot$ --
see Disney, et al. in these proceedings).  Additionally, the
impact parameter of LSB galaxies range from  less than 1 kpc through
well over 100 kpc at the 10$^{18}$ cm$^{-2}$ limit.

The results of the above properties, combined with the survey results
of the HIPASS equatorial survey and the massive LSB galaxy surveys
of Schombert, et.al (2002) and O'Neil \& Bothun (2001) indicate the
number of massive (M$_{HI}$ $>$ 10$^{10}$ M$_\odot$) LSB galaxies
is far higher than was previously thought.
This implies that LSB galaxies could be contributing far more to the z $<$ 0.1
gas mass density and HI luminosity function than is often
believed. 

Finally, in regards to Lyman-$\alpha$ absorption systems, the recent
observations regarding the mass, extent, and number density of 
LSB galaxies make it highly likely that LSB galaxies are important
contributors to Lyman-$\alpha$ absorption systems, both the damped Lyman-$\alpha$
systems and the Lyman-$\alpha$ absorbers at lower column density.


\clearpage
\begin{references}
\reference{Bell, E.F., et al. 2000 \mnras\ 312, 470}
\reference{Bijersbergen, M., de Blok, W. J. G., \& van der Hulst, J. M. 1999 \aa\ 351, 903}
\reference{Bothun, G. D., Schombert, J., Impey, C., Sprayberry, D., McGaugh, S. 1993 \aj\ 106, 530}
\reference{Bowen, D., Tripp\, T., \& Jenkins, E. 2001 \aj\ 121, 1456}
\reference{Chen, H-W,. Lanzetta, K. M., Webb, J. K., \& Barcons, X. 1998 \apj\ 498, 77}
\reference{de Blok, W. J. G., McGaugh, S. S., \& van der Hulst, J. M.  1996 \mnras\ 283, 18}
\reference{de Blok, W. J. G., van der Hulst, J. M., \& Bothun, G.D. 1995 \mnras\ 274, 235}
\reference{Evans, Rh., Davies, J. I., \& Phillipps, S.  1990 \mnras\ 245, 164}
\reference{Gerritsen,  P. \& de Blok, W.J.G 1999 A\&A 342, 655}
\reference{Impey, C., Bothun, G., \& Malin, D.  1988 \apj\ 330, 634}
\reference{Matthews, L., Gallagher, J., van Driel, W. 1999 \aj\ 118, 2751}
\reference{McGaugh, S., Schombert, J., \& Bothun G. 1995 \aj\ 109, 2019}
\reference{O'Neil, K., \& Bothun, G.D. 2001, in preparation}
\reference{O'Neil, K., Bothun, G.D., \& Schombert, J. 2000 \aj\ 119, 136}
\reference{O'Neil, K., McGaugh, S. S., \& Verheijen, M.A.W. 1999 \aj\ 119, 2194}
\reference{O'Neil, K., Bothun, G.D., Schombert, J., Impey, C.D. \&  McGaugh, S.
1998 \aj\ 116, 657}
\reference{O'Neil, K., Bothun, G.D., Cornell, M. 1997a \aj\ 113, 1212}
\reference{O'Neil, K., Bothun, G.D., Schombert, J., Cornell, M., \& Impey, C. 1997b \aj\ 114, 2448}
\reference{Pickering, T. E., Impey, C. D., van Gorkom, J. H., \& Bothun, G. D.  1997 \aj\ 114, 1858}
\reference{Rao, S.M., \& Turnshek, D.A. 1998 \apj\ 500, L115}
\reference{Rauch, M., Weyman, R.J., \& Morris, S.L. 1996 \apj\ 458, 518}
\reference{Schombert, J., O'Neil, K., \& Eder, J. 2002, in preparation}
\reference{Schombert, J.M.  1998 \aj\ 116, 1650}
\reference{Steidel, C., Pettini, M., Dickinson, M., \& Persson, S. 1994 \aj\ 108, 2046}
\reference{Turnshek, D., et al. 2001 \apj\ 553, 288}
\reference{Zwaan, M.A., Verheijen, M.A.W., Briggs, F. 1999 PASA 16, 100}
\end{references}
\end{document}